\documentstyle[preprint,pre,aps]{revtex}
\begin{document}
\draft
\title{Symmetry and dynamics of a magnetic oscillator
}

\author{
        Sang-Yoon Kim
        \footnote{Electronic address: sykim@cc.kangwon.ac.kr}
       }
\address{
Department of Physics\\ Kangwon National University\\
Chunchon, Kangwon-Do 200-701, Korea
}
\maketitle

\begin{abstract}
We consider a permanent magnetic dipole in an oscillating
magnetic field. This magnetic oscillator has two dynamical symmetries.
With increasing the amplitude $A$ of the magnetic field, dynamical
behaviors associated with the symmetries are investigated.
For small $A$, there exist symmetric states with respect to one of the
two symmetries. However, such symmetric states lose their symmetries
via symmetry-breaking pitchfork bifurcations and then the
symmetry-broken states exhibit period-doubling transitions to chaos.
Consequently, small chaotic attractors with broken symmetries appear.
However, as $A$ is further increased they merge into a large symmetric
chaotic attractor via symmetry-restoring attractor-merging crisis.
\end{abstract}


\narrowtext

 We consider a magnetic oscillator, consisting of a permanent
magnetic dipole of moment $m$ placed in a spatially uniform
magnetic field $B$ that oscillates periodically in time. Its motion
can be described by a second-order nonautonomous ordinary
differential equation \cite{Croquette,Schmidt1,Briggs},
\begin{equation}
I {\ddot \theta} + b {\dot  \theta} + m B_{ac} \cos \omega t
\sin \theta =0,
\label{eq:MP1}
\end{equation}
where the overdot denotes the differentiation with respect to time,
$\theta$ is the angle between the magnetic dipole and the magnetic
field, $I$ is the momemt of inertia of the magnetic dipole about a
rotation axis, $b$ is the damping parameter, and $B_{ac}$ and
$\omega$ are the amplitude and frequency of the periodically
oscillating magnetic field $B$, respectively.
Making the normalization
$\omega t \rightarrow 2 \pi (t+{1 \over 2})$ and $ \theta \rightarrow
2 \pi x$, we have
\begin{equation}
 {\ddot x} + \Gamma {\dot  x} - A \cos 2 \pi t \sin 2 \pi x =0,
\label{eq:MP2}
\end{equation}
where $x$ is a normalized angle with range
$x \in [-{1 \over 2}, {1 \over 2})$,
$\Gamma={ {2 \pi b} \over {I \omega}}$ and
$A = { {2 \pi m B_{ac}} \over {I \omega^2}}$.
Note also that this equation of motion is the
same as that of a particle in a standing wave field
\cite{Bialek,Schmidt2}.

 For the conservative case of $\Gamma=0$, the Hamiltonian system
exhibits period-doubling bifurcations and large-scale stochasticity
as the normalized amplitude $A$ is increased, which have been found
both numerically \cite{Bialek,Schmidt2} and experimentally
\cite{Croquette,Briggs}. Here we are interested in the dissipative
case of $\Gamma \neq 0$. An experiment of period-doubling bifurcations
in a dissipative system has been reported \cite{Schmidt1}.

 The normalized equation of motion (\ref{eq:MP2}) is reduced to two
first-order ordinary differential equations:
\begin{mathletters}
\begin{eqnarray}
{\dot x} &=& y,  \\
{\dot y} &=& - \Gamma y + A \cos 2 \pi t \sin 2 \pi x.
\end{eqnarray}
\label{eq:MP3}
\end{mathletters}
These equations have two symmetries $S_1$ and $S_2$, because
the transformations
\begin{eqnarray}
S_1: && x \rightarrow x \pm {1 \over 2},~y \rightarrow y,
~t \rightarrow t \pm {1 \over 2}, \label{eq:S1} \\
S_2: && x \rightarrow -x,~y \rightarrow -y,~ t  \rightarrow t,
\label{eq:S2}
\end{eqnarray}
leave Eq.~(\ref{eq:MP3}) invariant.
The transformations in Eqs.~(\ref{eq:S1}) and (\ref{eq:S2}) are
just the shift in both $x$ and $t$ and the inversion, respectively.
Hereafter we will call $S_1$ and $S_2$ the shift and inversion
symmetries, respectively. If an orbit $z(t) [\equiv (x(y),y(t))]$
is invariant under $S_i$ $(i=1,2)$, it is called an
$S_i$-symmetric orbit. Otherwise, it is called an $S_i$-asymmetric
orbit and has its ``conjugate'' orbit $S_i z(t)$.

 In this paper, with increasing the amplitude $A$ we study the
evolutions of both the stationary states and the rotational states
of period $1$ in the magnetic oscillator for a
moderately damped case of $\Gamma=1.38$. Dynamical behaviors
associated with the symmetries are particularly investigated.
As will be seen below, their dynamical symmetries are eventually
broken through symmetry-breaking pitchfork bifurcations, which
result in the birth of completely symmetry-broken states. These
symmetry-broken states undergo period-doubling transitions to chaos,
leading to small chaotic attractors with broken symmetries. However,
with further increasing $A$ they merge into a large symmetric
chaotic attractor through a symmetry-restoring attractor-merging
crisis \cite{Crisis}.

The surface of section for the periodically-driven magnetic oscillator
is the Poincar\'{e} time-$1$ map. Hence the Poincar\'{e} maps of an
initial point $z_0$ $[=(x_0,y_0)]$ can be computed by sampling the
orbit points $z_m$ at discrete time $t=m$ $(m=1,2,3, \dots)$. We call
the transformation $z_m \rightarrow z_{m+1}$ the Poincar\'{e} map
and write $z_{m+1}= P (z_m)$.

The linear stability of a $q$-periodic orbit of $P$ such that
$P^q(z_0) = z_0$ is determined from the linearized-map matrix
$DP^q(z_0)$ of $P^q$ at an orbit point $z_0$.
Here $P^q$ means the $q$-times iterated map.
Using the Floquet theory \cite{Gukenheimer1}, the matrix
$M$ $(\equiv DP^q)$ can be obtained by integrating the linearized
equations for small displacements,
\begin{mathletters}
\begin{eqnarray}
  \delta {\dot x} &=& \delta y,  \\
  \delta {\dot y} &=& - \Gamma \delta y + 2 \pi A \cos 2 \pi t
  \cos 2 \pi x \, \delta x
\end{eqnarray}
\label{eq:LE}
\end{mathletters}
with two initial displacements $(\delta x, \delta y) = (1,0)$ and
$(0,1)$ over the period $q$. The eigenvalues, $\lambda_1$ and
$\lambda_2$, of $M$ are called the Floquet (stability) multipliers,
characterizing the orbit stability. After some algebra, we find
that the determinant of $M$ is given by $det(M)= e^{-\Gamma q}$.
Hence the pair of Floquet multipliers of a periodic orbit lies either
on the circle  of radius  $e^{-\Gamma q/2}$ or on the  real axis
in the complex plane. The periodic orbit is stable  only when both
Floquet multipliers lie inside the unit circle.
Hence the periodic orbit can lose its stability only when a
Floquet multiplier $\lambda$ decreases (increases) through $-1$
$(1)$ on the real axis. When a Floquet multiplier $\lambda$ decreases
through $-1$, the periodic orbit loses its stability via
period-doubling bifurcation. On the other hand, when a Floquet
multiplier $\lambda$ increases through $1$, it becomes unstable via
pitchfork or saddle-node bifurcation.
For more details on bifurcations, refer to Ref.~\cite{Gukenheimer2}.

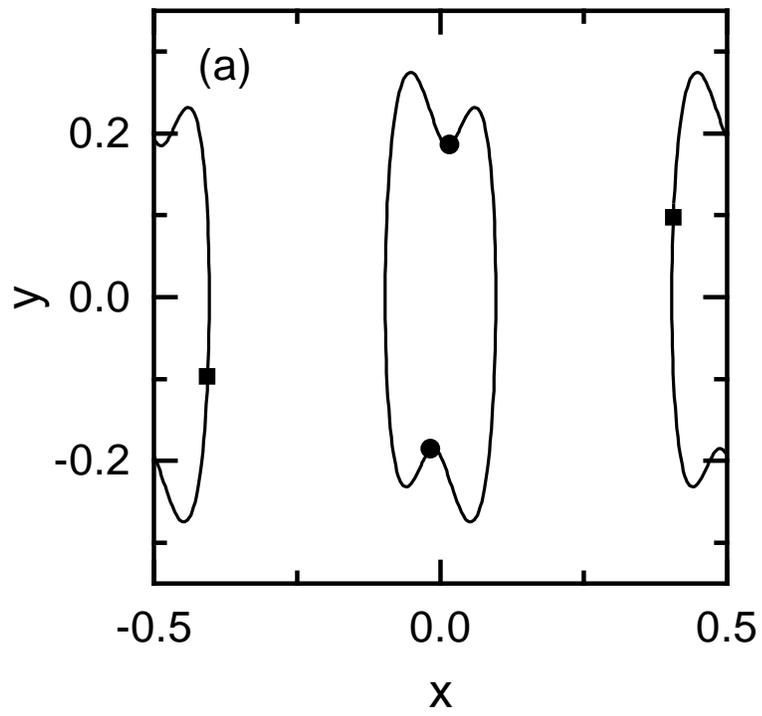
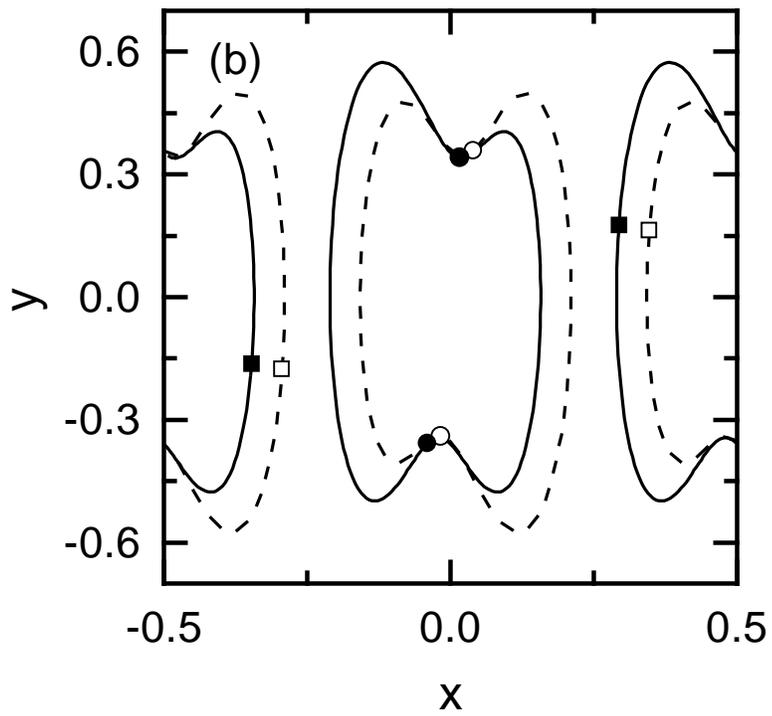
\begin{figure}
\caption{Symmetry-conserving and symmetry-breaking bifurcations.
         Two stable period-doubled orbits with the inversion
         symmetry $S_2$ are born from the two stationary points
         via symmetry-conserving period-doubling bifurcations.
         They are shown in (a) for $A=3.31$. Their phase flows
         (Poincar\'{e} maps) are denoted by solid lines [solid
         symbols (circle and square)]. The two $S_2$-symmetric
         orbits with period $2$ become unstable via $S_2$-symmetry
         breaking pitchfork bifurcations. Consequently, two
         $S_2$-conjugate pairs of stable orbits with period $2$
         appear, as shown in (b) for $A=3.87$. For each
         $S_2$-conjugate pair, the phase
         flow (Poincar\'{e} map) of one orbit is denoted by a solid
         line [solid symbol (circle or square)], while that of the
         other one is represented by a(n) dashed line [open symbol
         (circle or square)].
        }
\label{fig:ssb}
\end{figure}

 We first consider the case of the stationary states. The magnetic
oscillator has two stationary states $\hat{z}$'s. The first one is
$\hat{z}_I$ $=(0,0)$ and the second one is $\hat{z}_{II}$
$=({1 \over 2},0)$. These stationary states are symmetric ones with
respect to the inversion symmetry $S_2$, while they are asymmetric
and conjugate ones with respect to the shift symmetry $S_1$.
Hence they are partially symmetric orbits with only the inversion
symmetry $S_2$. We also note that the two stationary states
are the fixed points of $P$ [i.e., $P(\hat{z})=\hat{z}$
$(\hat{z}= \hat{z}_I,\, \hat{z}_{II})$]. With increasing $A$,
we investigate the evolutions of the fixed points
$\hat{z}_I$ and $\hat{z}_{II}$. For $A=3.142\,710\, \cdots$,
each fixed point loses its stability through a symmetry-conserving
period-doubling bifurcation, leading to the birth of a new stable
$S_2$-symmetric orbit with period $2$. An example for $A=3.31$ is
shown in Fig.~\ref{fig:ssb}(a). Like the stationary points, the
two stable period-doubled orbits with the inversion symmetry $S_2$
, whose phase flows are denoted by solid lines, are asymmetric and
conjugate ones with respect to the shift symmetry $S_1$.
The Poincar\'{e} map of the stable 2-periodic orbit
encircling the unstable fixed point $\hat{z}_I$ [$\hat{z}_{II}$] is
also represented by a solid circle (square).
However, as $A$ is further increased, each of
the two $S_2$-symmetric orbits of period $2$ becomes unstable
via $S_2$-symmetry breaking pitchfork bifurcation for
$A=A_{b,s}$ $(=3.817\,897\, \cdots).$ Consequently, two conjugate
pairs of new $S_2$-symmetry broken orbits with period $2$ appear for
$A> A_{b,s}$. An example for $A=3.87$ is given in
Fig.~\ref{fig:ssb}(b). One $S_2$-conjugate pair encircles the
unstable $\hat{z}_I$, while the other pair encircles the unstable
$\hat{z}_{II}$. For each $S_2$-conjugate pair, the phase flow
(Poincar\'{e} map) of one orbit is denoted by a solid line
[solid symbol (circle or square)], whereas that of the other one is
represented by a(n) dashed line [open symbol (circle or square)].
Thus the two symmetries $S_1$ and $S_2$ are completely broken.

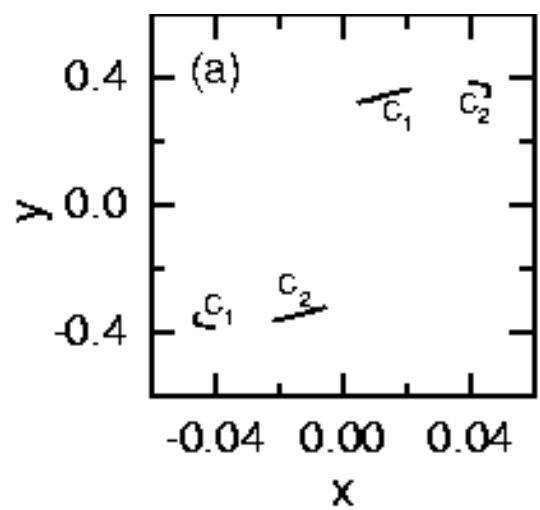
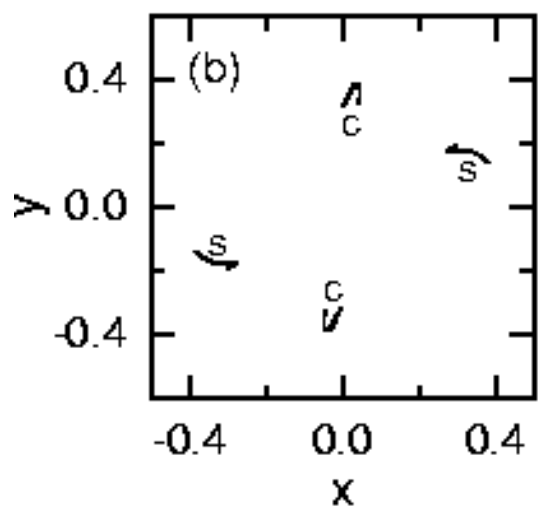
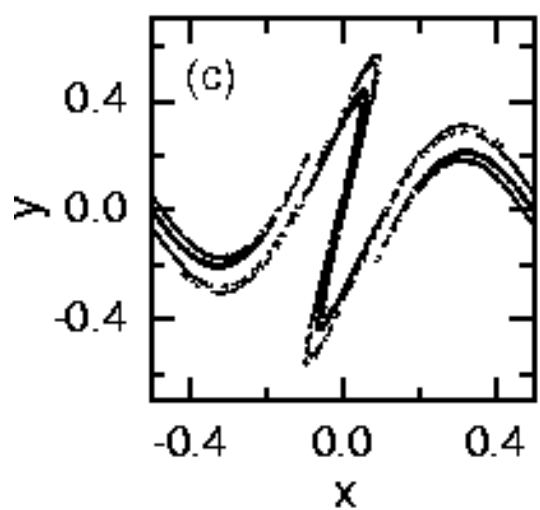
\begin{figure}
\caption{Poincar\'{e}-map plots of chaotic attractors after the
         period-doubling transitions to chaos of the stationary
         states. An $S_2$-conjugate pair of small chaotic attractors
         $c_1$ and $c_2$ near the unstable $\hat{z}_I$ are shown in
         (a) for $A=3.9411$. Each of them consists of two pieces.
         The two chaotic attractors $c_1$ and $c_2$ merge into
         a bigger one $c$ via $S_2$-symmetry restoring crisis.
         For $A=3.95$ the chaotic attractor $c$ with the inversion
         symmetry $S_2$ and its $S_1$-conjugate one, denoted by $s$,
         are shown in (b).
         The two chaotic attractors $c$ and $s$ also merge into a
         larger one via $S_1$-symmetry restoring crisis. A single
         large chaotic attractor with the completely restored $S_1$
         and $S_2$ symmetries is shown in (c) for $A=3.975$.
        }
\label{fig:ssr}
\end{figure}

With further increasing $A$, each of the four $2$-periodic
orbits with completely broken symmetries undergoes an
infinite sequence of period-doubling bifurcations, ending at a
finite critical point $A^*_s$ $(=3.934\,787\, \cdots)$ as in the
case of the one-dimensional maps \cite{Feigenbaum}.
For $A>A^*_s$, four small $S_1$- and $S_2$-asymmetric chaotic
attractors with positive largest Lyapunov exponent $\sigma$,
characterizing the average exponential rate of divergence
of nearby orbits \cite{Lexp}, appear. As $A$ is further increased
the different parts of each chaotic attractor coalesce and form
larger pieces. Through such a band-merging process, each chaotic
attractor eventually becomes composed of two pieces. An example for
$A=3.9411$ is given in Fig.~\ref{fig:ssr}(a). For the sake of
convenience, only two chaotic attractors with $\sigma \simeq 0.189$,
denoted by $c_1$ and $c_2$, near the unstable ${\hat z}_I$ are shown;
in fact, their conjugate chaotic attractors with respect to the $S_1$
symmetry exist near the unstable ${\hat z}_{II}$. As $A$ exceeds a 1st
critical value $A_{c,1}$ $(=3.9484)$, the two chaotic attractors $c_1$
and $c_2$ merge into a bigger one $c$ via $S_2$-symmetry restoring
crisis. As an example, a chaotic attractor $c$ with
$\sigma \simeq 0.307$ is shown in Fig.~\ref{fig:ssr}(b) for $A=3.95$,
and its conjugate one with respect to the $S_1$ symmetry is also
denoted by $s$. These two chaotic attractors $c$ and $s$ become
$S_2$-symmetric (but still $S_1$-asymmetric) ones. Thus the inversion
symmetry $S_2$ is first restored. However, as $A$
passes through a 2nd critical value $A_{c,2}$ $(=3.9672)$ the two
small chaotic attractors, $c$ and $s$, also merge to form a larger one
via $S_1$-symmetry restoring crisis. An example for $A=3.975$ is
shown in Fig.~\ref{fig:ssr}(c). Note that the single large chaotic
attractor with $\sigma \simeq 0.599$ is both $S_1$- and
$S_2$-symmetric one. Thus the two symmetries $S_1$ and $S_2$ are
restored completely, one by one via two symmetry-restoring crises.

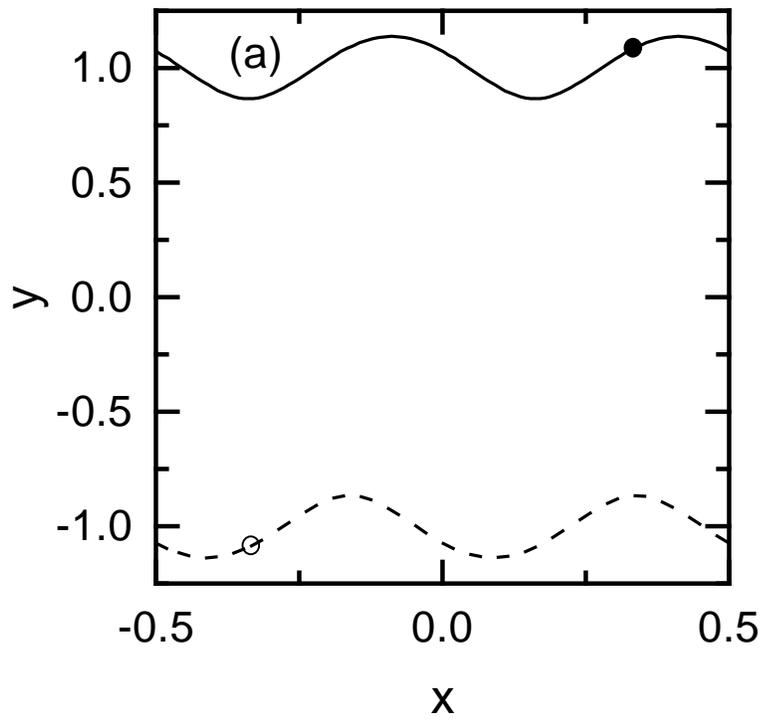
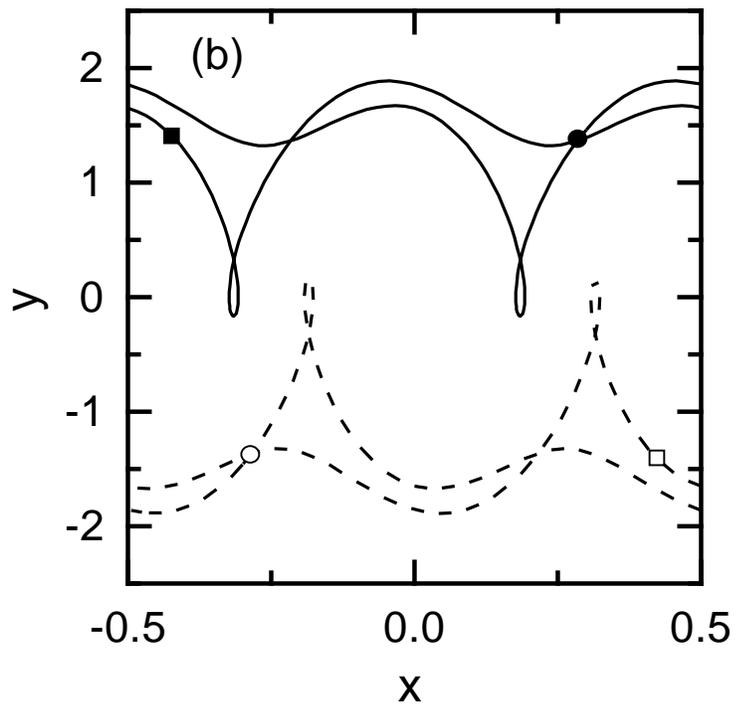
\begin{figure}
\caption{Saddle-node and symmetry-breaking pitchfork bifurcations.
         A conjugate pair of the $S_2$-asymmetric rotational orbits
         with period $1$ born via saddle-node bifurcations is shown in
         (a) for $A=3.31$. These two orbits of period $1$ are
         $S_1$-symmetric ones. The phase flow (Poincar\'{e} map) of
         the orbit with positive angular velocity is denoted by a
         solid line (solid circle), whereas that of the other orbit
         with negative angular velocity is represented by a(n) dashed
         line (open circle). The two $S_1$-symmetric orbits with
         period $1$ lose their stability via $S_1$-symmetry breaking
         pitchfork bifurcations. Consequently, two $S_1$-conjugate
         pairs of stable orbits with period $1$ appear, as shown in
         (b) for $A=11.1$. The phase flow (Poincar\'{e} map) of one
         $S_1$-conjugate pair is denoted by a solid line [solid
         symbols (circle and square)], whereas that of the other pair
         is represented by a(n) dashed line [open symbols (circle and
         square)].
        }
\label{fig:rsb}
\end{figure}

 We now study the evolution of the rotational states of period $1$ by
increasing $A$. A pair of stable and unstable rotational orbits with
period $1$ is born for $A \simeq 2.771$ through a saddle-node
bifurcation. In contrast to the stationary states, the rotational
states are $S_1$-symmetric, but $S_2$-asymmetric ones. As an example,
a conjugate pair of $S_2$-asymmetric rotational states for $A=3.31$
is shown in Fig.~\ref{fig:rsb}(a). The phase flow (Poincar\'{e} map)
of the orbit with positive angular velocity is denoted by a solid
line (solid circle), while that of the other orbit with negative
angular velocity is represented by a(n) dashed line (open circle).
With increasing $A$ further, each
stable $S_1$-symmetric rotational orbit with period $1$ loses
its stability for $A=A_{b,r}$ $(=9.892\,445\, \cdots)$ via
$S_1$-symmetry breaking pitchfork bifurcation.
Consequently, two conjugate pairs of $S_1$-symmetry broken orbits
with period $1$ appear for $A>A_{b,r}$.
An example for $A=11.1$ is given in Fig.~\ref{fig:rsb}(b).
The phase flow (Poincar\'{e} map) of one $S_1$-conjugate pair
with positive average angular velocity is
denoted by a solid line [solid symbols (circle and square)], while
that of the other pair with negative average angular velocity
is represented by a(n) dashed line [open
symbols (circle and square)].
Thus the two symmetries $S_1$ and $S_2$ are completely broken.

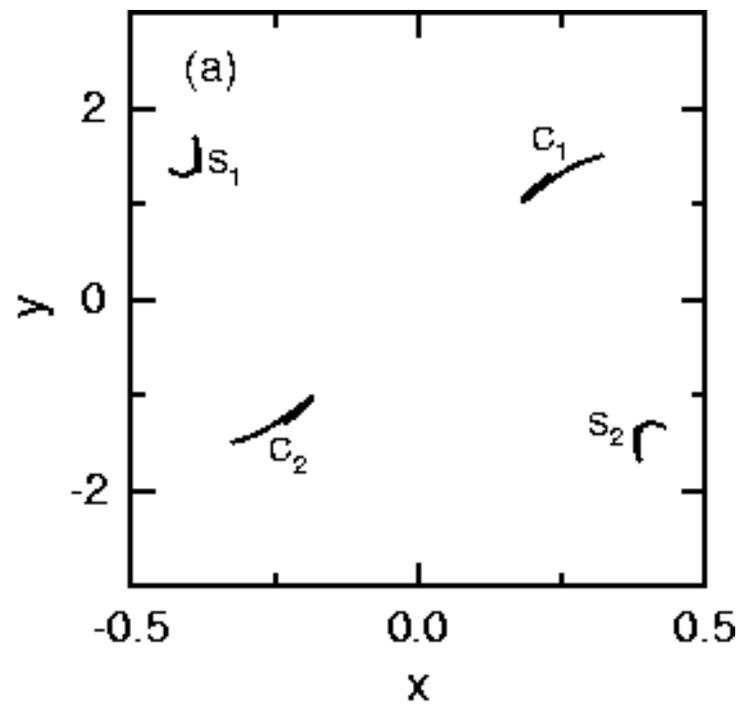
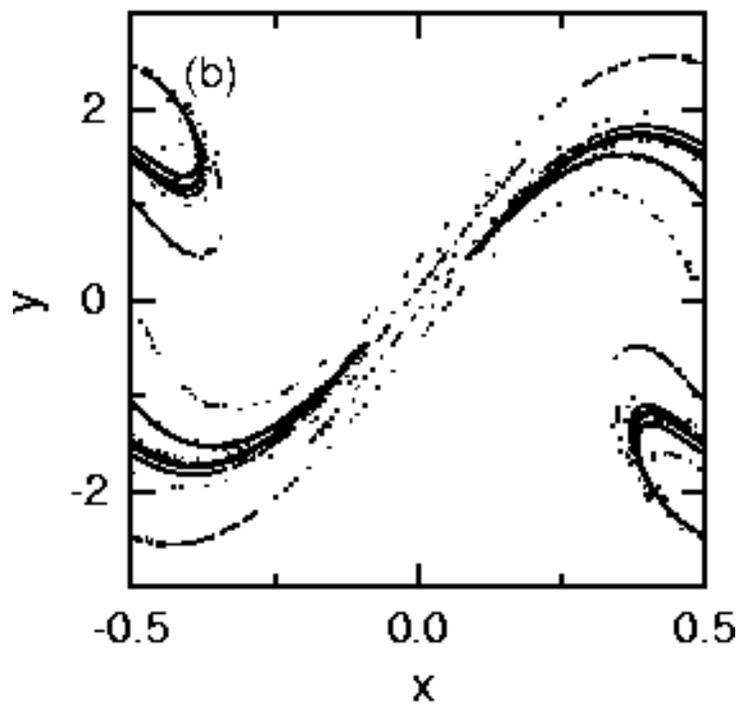
\begin{figure}
\caption{Poincar\'{e}-map plots of chaotic attractors after
         the period-doubling transitions to chaos of the rotational
         states with period $1$. Four small chaotic attractors,
         denoted by $c_1$, $c_2$, $s_1$, and $s_2$, are shown in (a)
         for $A=12.342$. These four small chaotic attractors merge
         into a larger one via $S_1$- and $S_2$-symmetries restoring
         crisis. A single large chaotic attractor with the
         simaltaneously resotred $S_1$ and $S_2$ symmetries is shown
         in (b) for $A=12.4$.
        }
\label{fig:rsr}
\end{figure}

With further increase of $A$, each of the four $1$-periodic rotational
orbits with completely broken symmetries exhibits an infinite
sequence of period-doubling bifurcations, accumulating at a finite
critical point $A^*_r$ $(=12.252\,903\, \cdots)$, as in the case of
the stationary states. For $A>A^*_r$, four small chaotic attractors
appear. Through a band-merging process, the different parts of a
chaotic attractor merge into larger pieces. Thus each chaotic
attractor eventually consists of a single piece, as illustrated
in Fig.~\ref{fig:rsr}(a) for $A=12.342$. Four small chaotic attractors
with $\sigma \simeq 0.416$ are denoted by $c_1$, $c_2$, $s_1$, and
$s_2$, respectively. However, as $A$ passes through a critical value
$A_c$ $(=12.3424)$, the four small chaotic attractors merge to form a
larger one via $S_1$- and $S_2$-symmetries restoring crisis.
An example for $A=12.4$ is given in Fig.~\ref{fig:rsr}(b). We note
that the single large chaotic attractor with $\sigma \simeq 0.713$ has
both the $S_1$ and $S_2$
symmetries. Thus the two symmetries $S_1$ and $S_2$ are restored
simultaneously through one symmetry-restoring crisis, which is in
contrast to the case of the stationary states.

\acknowledgements
This work was supported by the Basic Science Research Institute
Program, Ministry of Education, Korea, Project No. BSRI-97-2401.

\end{document}